\documentclass{article}

\usepackage{spconf,graphicx}
\usepackage{amsmath,amssymb,amsfonts}
\usepackage{svg}
\usepackage{booktabs}
\usepackage[colorlinks=true, allcolors=black]{hyperref}

\title{INTEGRATION OF CALCIUM IMAGING TRACES VIA DEEP GENERATIVE MODELING}

\name{B. Ros$^{\star}$ \quad M. Olives-Verger$^{\dagger}$\quad C. Fuses$^{\star}$\quad JM. Canals$^{\star}$\quad J. Soriano$^{\dagger}$ \quad J. Abante$^{\star}$\thanks{Corresponding author: \texttt{jordi.abante@ub.edu}}}
  
\address{$^{\star}$ Dept. of Biomedical Sciences \& Institute of Neurosciences, Universitat de Barcelona, Barcelona, Spain \\
$^{\dagger}$ Dept. of Condensed Matter Physics, Universitat de Barcelona, Barcelona, Spain}

\begin{document}
%
\maketitle
\begin{abstract}
Calcium imaging allows for the parallel measurement of large neuronal populations in a spatially resolved and minimally invasive manner, and has become a gold-standard for neuronal functionality. While deep generative models have been successfully applied to study the activity of neuronal ensembles, their potential for learning single-neuron representations from calcium imaging fluorescence traces remains largely unexplored, and batch effects remain an important hurdle. To address this, we explore supervised variational autoencoder architectures that learn compact representations of individual neurons from fluorescent traces without relying on spike inference algorithms. We find that this approach outperforms state-of-the-art models, preserving biological variability while mitigating batch effects. Across simulated and experimental datasets, this framework enables robust visualization, clustering, and interpretation of single-neuron dynamics.
\end{abstract}

\vspace{-10pt}

\section{Introduction}
\label{sec:intro}
Calcium imaging (Ca$^{2+}$Im) of neuronal activity has emerged as a powerful alternative to electrophysiology, allowing minimally invasive, spatially resolved recordings from large neuronal populations~\cite{kerr2008imaging}. In contrast to electrophysiology, this technique facilitates the study of the interplay between neuron location and activity, e.g., in the context of complex systems~\cite{soriano2023neuronal} or neuroengineering~\cite{yamamoto2018,montalaflaquer2022}.

Ca$^{2+}$Im data has been extensively used to produce neuronal population models that facilitate the study of the neuronal origin of behavior. Recently, deep generative models (DGMs) have been proposed to model neuronal population dynamics as a function of behavioral queues ~\cite{Pandarinath2018,Zhou2020,Schneider2023}. These models attempt to find a latent representation $\mathbf{Z}(t)$ of the population state $\mathbf{X}(t)$, producing meaningful lower-dimensional neural population dynamics~\cite{Zhou2020,Jazayeri2021} beyond the limitations of linear methods.
This work has had a profound impact in neuroscience, providing new powerful approaches for visualization, clustering, or discovery of latent spaces that explain neuronal variance that facilitate the prediction of a future behavior based on past neural activity~\cite{Schneider2023}. These approaches, however, focus on finding useful representations for the population state at a given time.

Here, our primary objective is to address the challenge of integrating fluorescent traces into a common latent space that captures the underlying firing dynamics of neurons. Traditional methods, such as PCA~\cite{Comella-Bolla2020}, focus on reducing dimensionality at the level of single-neuron representations, but this results in confounding technical and biological variability. In addition, existing approaches often rely on spike inference~\cite{Friedrich2017,Speiser2017}, which introduces sensitivity to model assumptions and can obscure the true neural activity patterns. By leveraging deep generative models (DGMs), we aim to learn a latent representation of single-cell Ca$^{2+}$Im where disparate fluorescent signals are coherently mapped, thereby facilitating downstream tasks like visualization and clustering of individual neurons~\cite{murphy2023}. In particular, we explore the potential of supervised variational autoencoders (SVAEs) to produce biologically meaningful latent representations while minimizing batch effects. This approach is able to isolate biological variability, eliminating confounding effects and revealing the true dynamical behavior of neurons. To introduce an inductive bias capturing the temporal structure inherent to Ca$^{2+}$Im traces, we extend implement a version of SVAE with a Gaussian Process (GP) likelihood in the decoder (GPVAE). Unlike approaches that impose temporal priors on the latent space~\cite{casale2018gaussian}, our formulation directly models temporal correlations in the observations while maintaining single-cell resolution, enhancing the interpretability of learned representations. In simulations, SVAEs models achieve a favorable balance between batch-effect correction and retention of biological variability. In addition, we use this framework to integrate multispecies Ca$^{2+}$Im data, including mouse and rat data, showing fundamental differences in firing dynamics between species. Altogether, our results show that this framework produces superior embeddings which, in turn, facilitates downstream analysis.

\vspace{-15pt}

\section{Methodology}
\label{sec:methods}

\subsection{Synthetic data}
We generate realistic calcium traces simulations to evaluate the quality of learned representations. Neuronal networks are constructed with spatially constrained connectivity and realistic axonal growth mechanisms~\cite{orlandi2013,Houben2025}. From each network, we define a weighted adjacency matrix $\mathbf{W}$, with excitatory and inhibitory connection strengths $w_E = 2$ and $w_I = -0.5$, respectively, and inhibitory neurons are randomly assigned (60-80\%). Neuronal dynamics follow Izhikevich’s integrate-and-fire model~\cite{izhikevich}:
\begin{equation}
\begin{aligned}
\frac{dv}{dt} &= 0.04 v^2 + 5 v + 140 - u + I + \eta, \
\frac{du}{dt} &= a (b v - u)~,
\label{eq:izhi}
\end{aligned}
\end{equation}
where $v$ is the membrane potential, $u$ the recovery variable, $I$ the summed synaptic input from $\mathbf{W}$, and $\eta$ a noise term. A neuron spike occurs when $v$ reaches a threshold of $30$~mV, followed by resetting $v \leftarrow v_0$ and $u \leftarrow u + \Delta u$. The parameters $a, b$, $v_0$, and $\Delta u$ can be adjusted to account for different neuronal firing behaviors~\cite{izhikevich}.

The resulting binary spikes are transformed into calcium fluorescence traces $\mathbf{f}$ using a biophysical model~\cite{Deneux2016}:
\begin{equation}
f_t = F_0 * A (\kappa + p_2 (\kappa^2 - \kappa) + p_3 (\kappa^3 - \kappa)),~~t=1,\ldots,T,
\end{equation}
with $F_0$ baseline, $A$ amplitude, and $(p_2, p_3)$ fitted to mimic GCaMP6s~\cite{wu2019}. Afterwards, Gaussian noise with standard deviation $\sigma_t$ and min–max normalization are applied to mimic experimental recordings. 

We simulate several synthetic cultures with $N = 1000$ neurons each from 2 distinct groups with different proportions of neuronal dynamics: Group 1 (20\% RS, 80\% LTS) and Group 2 (40\% RS, 60\% LTS) (Fig.~\ref{fig:fig_1}A). Different dynamics are selected by modifying parameters $a$, $b$, $v_0$, and $\Delta u$ in Eq.~(\ref{eq:izhi}) (Table~\ref{tab:neuron_parameters}). Three samples per group are generated with shared connectivity and small random parameter variations.

\begin{table}[h]
    \centering
    \begin{tabular}{lcccc}
        \toprule
        \textbf{Neuron Type} & \textbf{$a$} & \textbf{$b$} & \textbf{$v_0$} & \textbf{$\Delta u$} \\
        \midrule
        Regular spiking (RS) & 0.02 & 0.2 & -65 & 8 \\
        Low-threshold spiking (LTS) & 0.02 & 0.25 & -65 & 2 \\
        \bottomrule
    \end{tabular}
    \vspace{3pt}
    \caption{Parameters for different neuron types.}
    \label{tab:neuron_parameters}
    \vspace{-20pt}
\end{table}

\subsection{Experimental data}
We analyze spontaneous activity recordings from neuronal cultures prepared from primary cortical neurons of rat and mouse embryos (E16–E18). Neurons are cultured from rat and mouse embryos under standardized conditions and transduced with the genetically encoded calcium indicator GCaMP6s, enabling fluorescence imaging of intracellular calcium dynamics linked to neuronal firing.
For each species, six independent batches of neuronal cultures are prepared --- three at day {\em in vitro} (DIV) 7 and three at DIV 12 --- under standardized conditions, each 6~mm in diameter and containing about 2000 neurons. Data is acquired on a microscope equipped for fluorescence together with a high-speed camera that provides images at 33 Hz and cellular resolution. 
Spontaneous neuronal activity is recorded for 10 min, and images processed with the software Netcal~\cite{orlandi2017netcal} to extract the fluorescence trace $f_i(t)$ of each neuron $i$. Traces are normalized as $\text{DFF}_i(t) \equiv [f_i(t) - f_{i,0}(t)]/f_{i,0}$, where $f_{i,0}$ is the fluorescence signal of neuron $i$ at rest, and subsequently normalized within each recording session.


\vspace{-10pt}
\subsection{Models}
\subsubsection{Bayesian Factor Analysis}
We first consider a linear generative model that extends principal component analysis (PCA) by modeling uncertainty in both the latent space and observation noise. Each calcium trace \( \mathbf{x}_n \in \mathbb{R}^D \) is generated from a latent variable \( \mathbf{z}_n \in \mathbb{R}^K \), drawn from a standard multivariate Gaussian prior: $\mathbf{z}_n \sim \mathcal{N}(\mathbf{0}, \mathbf{I})$. Conditioned on \( \mathbf{z}_n \), the observation is modeled as $\mathbf{x}_n \mid \mathbf{z}_n \sim \mathcal{N}(W \mathbf{z}_n + \boldsymbol{\mu}, \boldsymbol{\Psi})$, where \( W \in \mathbb{R}^{D \times K} \) is the factor loading matrix, \( \boldsymbol{\mu} \in \mathbb{R}^D \) is the mean vector, and \( \boldsymbol{\Psi} \in \mathbb{R}^{D \times D} \) is a diagonal covariance matrix capturing modality-specific noise. This models serves as the state of the art (SOTA) for dimensionality reduction of individual fluorescence traces obtained from Ca$^{2+}$Im.
\vspace{-10pt}
\subsubsection{Variational Autoencoder}
To capture non-linear structure in calcium traces, we implement a standard VAE with multilayer perceptrons (MLPs). Each trace $\mathbf{x}_n$ is encoded into a latent variable $\mathbf{z}_n \sim \mathcal{N}(\mathbf{0}, \mathbf{I})$. The decoder defines the likelihood over calcium trace \( \mathbf{x}_n \in \mathbb{R}^D \) as $\mathbf{x}_n \mid \mathbf{z}_n \sim \mathcal{N}(\boldsymbol{\mu}_\theta(\mathbf{z}_n), \sigma^2 \mathbf{I})$, where \( \boldsymbol{\mu}_\theta(\cdot) \) is a neural network parameterized by \( \theta \), and \( \sigma^2 \) is a fixed hyperparameter. Since the true posterior $p(\mathbf{z}_n \mid \mathbf{x}_n)$ is intractable, so we approximate it with a Gaussian variational posterior $q_\phi(\mathbf{z}_n \mid \mathbf{x}_n)$. The model is trained by maximizing the standard evidence lower bound (ELBO):
\begin{align*}
    \mathcal{L}_{\theta, \phi}(\mathbf{x}_n) 
    =& ~\mathbb{E}_{q_\phi(\mathbf{z}_n \mid \mathbf{x}_n)}\left[\log p_\theta(\mathbf{x}_n \mid \mathbf{z}_n)\right]\\
    &- \mathrm{KL}\left(q_\phi(\mathbf{z}_n \mid \mathbf{x}_n) \,\|\, p(\mathbf{z}_n)\right)~.
    \label{eq:elbo_unsup}
\end{align*}

\begin{figure*}[h]
    \centering
    \includegraphics[width=0.99\linewidth]{figures/figure_1_icassp_2026_v1.png}
    \caption{Simulation results. (A) Examples of simulated traces, consisting of two groups, with three replicates each, with varying proportions of neuronal dynamic behaviors (RSor LTS) shown in different colors. (B) Silhouette over firing labels vs. kBET score over batch labels for noise with $\sigma^2 = 1.0$, showing that the supervised models reduce the impact of the technical variability in the embedding, resulting in smaller values of kBET. (C) UMAP of latent representation for BFA (left) and SVAE (right) model for $K = 4$, showing the reduction of batch effect in the latent space with a supervised model. (D) kBET score over batch labels for each model over different levels of noise, showing that supervised models are more robust to noisy data.}
    \label{fig:fig_1}
\end{figure*}

\vspace{-10pt}
\subsubsection{Supervised Generative Model}
We extend the VAE framework to incorporate batch labels for modeling calcium traces \( \mathbf{x}_n \in \mathbb{R}^D \). Each trace is paired with a known batch label \( \mathbf{y}_n \in \{0, 1\}^B \), encoded as a one-hot vector across \( B \) batches. The generative process is defined as:
\begin{gather}
    \mathbf{y}_n\sim \mathrm{OneHotCategorical}\left(\tfrac{1}{B} \cdot \mathbf{1}\right)~, \\
    \quad
    \mathbf{z}_n \sim \mathcal{N}(\mathbf{0}, \mathbf{I})~, \\
    \quad
    \mathbf{x}_n \mid \mathbf{z}_n, \mathbf{y}_n \sim \mathcal{N}(\boldsymbol{\mu}_\theta(\mathbf{z}_n, \mathbf{y}_n), \sigma^2 \mathbf{I})~,
\end{gather}
where \( \boldsymbol{\mu}_\theta(\cdot, \cdot) \) is a decoder neural network that maps the latent code and label to the mean of the observation. The variational posterior is defined analogously as a Gaussian:
\begin{equation}
    q_\phi(\mathbf{z}_n \mid \mathbf{x}_n, \mathbf{y}_n) = \mathcal{N}(\boldsymbol{\mu}_\phi(\mathbf{x}_n, \mathbf{y}_n), \operatorname{diag}(\boldsymbol{\sigma}^2_\phi(\mathbf{x}_n, \mathbf{y}_n)))~,
\end{equation}
where \( \boldsymbol{\mu}_\phi \) and \( \boldsymbol{\sigma}^2_\phi \) are outputs of an encoder network that takes the concatenated vector \( [\mathbf{x}_n, \mathbf{y}_n] \in \mathbb{R}^{D + B} \) as input. The inclusion of the batch label \( \mathbf{y}_n \) allows the latent variable \( \mathbf{z}_n \) to focus on variability that is not class-specific. This disentangling enhances the interpretability and robustness of the learned latent space \cite{Siddharth2017}. This model is trained by maximizing the ELBO:

\begin{align*}
    \mathcal{L}_{\theta, \phi}(\mathbf{x}_n, \mathbf{y}_n) 
    =& ~\mathbb{E}_{q_\phi(\mathbf{z}_n \mid \mathbf{x}_n, \mathbf{y}_n)}\left[\log p_\theta(\mathbf{x}_n \mid \mathbf{z}_n, \mathbf{y}_n)\right]\\
    &- 
    \mathrm{KL}\left(q_\phi(\mathbf{z}_n \mid \mathbf{x}_n, \mathbf{y}_n) \,\|\, p(\mathbf{z}_n)\right)~.
    \label{eq:elbo_sup}
\end{align*}

\subsubsection{Gaussian Process Supervised Generative Model}
We extend the supervised generative model by incorporating a Gaussian Process (GP) in the decoder to explicitly capture temporal dependencies in calcium signals. Thus, in this case the likelihood now accounts for correlations defined via a kernel function. Specifically, we place a GP prior
\begin{equation}
   f_{\theta}(\cdot)\sim\mathcal{GP}~(m_\theta(\mathbf{z};\mathbf{y}), k_\theta(\mathbf{z}, \mathbf{z}')) 
\end{equation}
over $M$ evenly spaced inducing points $T_\mathbf{u}$, with mean function $m(\mathbf{z};\mathbf{y})$, and radial basis kernel
\begin{equation}
    k(\mathbf{z}, \mathbf{z}') = \ell^2 \exp\left({-\frac{\mid\mid \mathbf{z}-\mathbf{z}'\mid\mid^2}{2\sigma^2}}\right)~,
\end{equation}
where $\ell$ is the length-scale and $\sigma^2$ the variance, which will be optimized during training. 

The decoder predicts the GP mean values at the inducing points \(\boldsymbol{\mu}_\mathbf{u} = f_\theta(\mathbf{z}_n,\mathbf{y}_n)\in\mathbb{R}^\mathrm{M}\). Conditioning the GP on $\mu_\mathbf{u}$ yields the predictive mean and covariance on the full set of points $\boldsymbol{\mu}_\mathbf{f}$.
The reconstructed input is modeled as 
\begin{equation}
     \mathbf{x}_n \mid \mathbf{z}_n, \mathbf{y}_n
    \;\sim\; \mathcal{N}~(\boldsymbol{\mu}_\mathbf{f}, \text{diag}(\boldsymbol{\Sigma}_\mathbf{f}))~.
\end{equation}
Using only the diagonal of \(\boldsymbol{\mu}_\mathbf{f}\) improves tractability and stability, while the GP structure still enforces temporal smoothness through the predictive mean. Training proceeds by maximizing the ELBO, as in the SVAE.

\subsection{Evaluation Metrics}

To assess the quality of the learned latent representation $\mathbf{Z}$, we compute a series of complementary metrics that evaluate class separability, batch effect removal, clustering structure, and local neighborhood consistency. In order to measure the compactness and separation of clusters in latent space we compute the silhouette score, evaluated with respect to ground-truth neuronal firing dynamics. A high value indicates good clustering and separation of firing dynamics in latent space. To assess batch mixing we apply the kBET rejection rate~\cite{Bttner2019}, compares the local batch composition of each point's neighborhood to the global batch distribution using a chi-square test. A low rejection rate suggests that the representation is invariant to batch effects. All metrics were computed on a held-out test set that was not used during model training.

\section{Results}
\label{sec:results}
We evaluated the performance of supervised DGMs in integrating Ca$^{2+}$Im recordings from different batches by comparing it to baseline approaches like Bayesian factor analysis and a simple Variational Autoencoder.

\begin{figure}[t]
    \centering
    \includegraphics[width=\linewidth]{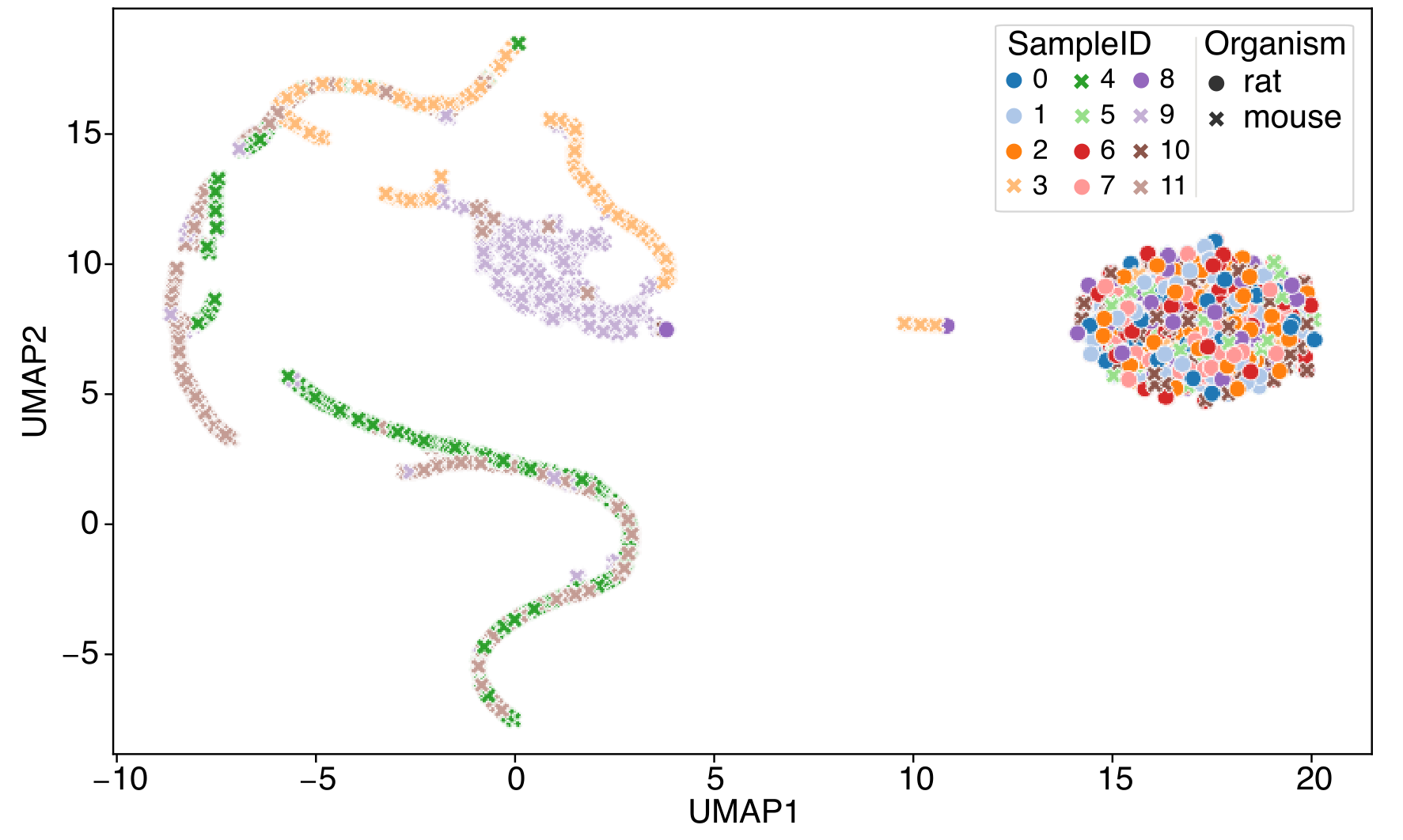}
    \caption{UMAP of experimental dataset obtained with SVAE for $K = 4$, showing the separation between mouse and rat traces and integration between experimental batches.}
    \label{fig:exp_UMAP}
\end{figure}

 We see that the addition of batch labels on the training set reduces batch effect by means of a lower kBET score, while maintaining an informative latent space (Fig.~\ref{fig:fig_1}B). This is consistent with the fact that the latent variable \(\bf Z\) is free to use its representational capacity to model characteristics of the signal that are not specific to a sample, since the variation between samples is provided by the labels \(\bf Y\) (Fig.~\ref{fig:fig_1}C). In contrast, the representation learned by both the BFA and VAE models is dominated by technical variability (Fig.~\ref{fig:fig_1}C). Overall, when assessing the quality of the latent representation, with respect to biological and technical variability, we observed that the supervised models offers a more favorable trade-off between preserving biological structure and removing recording session batch effects, while also beig more robust to noise (Fig.~\ref{fig:fig_1}D). We observe that as the number of latent dimensions increases, model performance tends to decline, indicated by higher kBET scores and lower silhouette scores. This suggests that the expanded latent space allows more capacity to capture batch-specific variability, resulting in the encoding of both biological differences between neurons and undesired batch effects.

Finally, for the experimental datasets we aim to separate species variability from batch effects. The UMAP visualization (Fig.~\ref{fig:exp_UMAP}) shows that the model effectively removes batch effects, grouping neurons by species rather than experimental batch. Specifically, neurons from rat cultures form a compact cluster, while those from mouse appear more dispersed in the latent space. This dispersion may suggest that mouse neurons are more sensitive to experimental conditions or that small variations in network development, e.g., due to neuronal density or biochemical signaling, affect neurons' behavior and their fluorescence signature. Two mouse culture batches (samples 5 and 10) cluster closer to rat neurons, suggesting shared fluorescence traits, such as sharp peaks from network-wide synchronous dynamics. This highlights abnormal development of the mouse cultures and may help identify outliers or motivate the exploration of the causes underlying abnormal behavior. 



\section{Discussion}
\label{sec:discussion}

Here we develop two supervised deep generative models for single-neuron Ca$^{2+}$Im data that leverage batch labels to reduce technical variability while preserving biologically meaningful signals. By removing batch effects from latent representations, the SVAE captures intrinsic neuronal dynamics more effectively than unsupervised approaches, and the GPVAE further models temporal correlations to improve reconstruction and interpretability. Across both synthetic and experimental datasets, these models consistently achieve lower kBET scores while maintaining informative latent spaces, demonstrating robustness to noise and technical variability. In real data, the SVAE model highlights atypical neuronal activity, enabling more precise investigation of heterogeneous neural populations. Future work should explore multimodal integration, combining Ca$^{2+}$Im with transcriptomic or immunohistochemistry labels, which would further facilitate interpretation.

\section{Conclusion}
\label{sec:conclusion}

We explore different SVAE architectures to integrate Ca$^{2+}$Im data. Compared with the state-of-the-art, the SVAE models provide robust latent representations to batch effects, and facilitate the identification of atypical neural activity in real data. Altogether, this framework represents a valuable tool for studying heterogeneous neuronal activity and can be extended for multimodal single-cell datasets.

\newpage
\bibliographystyle{IEEEbib}
\bibliography{refs}

\end{document}